\documentclass[twocolumn]{aastex63}

\newcommand{\cd}{d$^{-1}$}

\newcommand{\ms}{m\,s$^{-1}$}
\newcommand{\alpcir}{$\alpha$\,Cir}
\usepackage{upgreek}

\received{\ldots}
\revised{\ldots}
\accepted{\ldots}

\submitjournal{PASP}

\shorttitle{HRS observations of $\alpha$\,Circini}
\shortauthors{D.~L. Holdsworth \& E. Brunsden}

\begin{document}

\title{SALT HRS capabilities for time resolved pulsation analysis: a test with the roAp star $\alpha$\,Circini\footnote{based on observations made with the Southern African Large Telescope (SALT)}}

\correspondingauthor{D.~L. Holdsworth}
\email{dlholdsworth@uclan.ac.uk}

\author[0000-0003-2002-896X]{D.~L. Holdsworth}
\affiliation{Jeremiah Horrocks Institute, University of Central Lancashire, Preston PR1 2HE, UK}

\author[0000-0002-1555-1943]{E. Brunsden}
\affiliation{Department of Physics, University of York, Heslington, York, YO10 5DD, UK}

\begin{abstract}
Spectroscopy is a powerful tool for detecting variability in the rapidly oscillating Ap (roAp) stars. The technique requires short integrations times and high resolution, and so is limited to only a few telescopes and instruments. To test the capabilities of the High Resolution Spectrograph (HRS) at the Southern African Large Telescope (SALT) for the study of pulsations in roAp stars, we collected 2.45\,hr of high-resolution data of the well studied roAp star \alpcir\ in a previously unused instrument configuration. We extracted radial velocity measurements using different rare earth elements, and the core of H$_\alpha$, via the cross correlation method. We performed the same analysis with a set of \alpcir\ data collected with the High Accuracy Radial velocity Planet Searcher (HARPS) spectrograph to provide a benchmark for our SALT HRS test. We measured significant radial velocity variations in the HRS data and show that our results are in excellent agreement between the two data sets, with similar signal-to-noise ratio detections of the principal pulsation mode. With the HRS data, we report the detection of a second mode, showing the instrument is capable of detecting multiple and low-amplitude signals in a short observing window. We concluded that SALT HRS is well-suited for characterising pulsations in Ap stars, opening a new science window for the telescope. Although our analysis focused on roAp stars, the fundamental results are applicable to other areas of astrophysics where high temporal and spectral resolution observations are required.
\end{abstract}


\section{Introduction} 
\label{sec:intro}
The brightest member of the rapidly oscillating Ap (roAp) class of variable stars is $\alpha$\,Circini (HD\,128898; HR\,5463). Rapid oscillations were detected in the light curve of this star in 1981 \citep{kurtz1981}. Further observations were obtained and discussed in the seminal paper on the roAp stars \citep{kurtz1982}. It is thought that the roAp stars constitute just 4\,per\,cent of the Ap stars \citep{cunha2019}.

The Ap stars are a subset of the A stars which are characterised by their strong magnetic fields, chemical peculiarities, and generally slow rotation \citep[see][for a discussion on rotation in Ap stars]{2020A&A...639A..31M}. In the presence of a strong magnetic field, convection is suppressed which allows for the gravitational settling and radiative levitation of elements in the stellar atmosphere. In low-resolution classification spectra, the Ap stars are usually identified by absorption lines of Si, Sr, Eu or Cr, or a mixture of these \citep[e.g.,][]{maury1897,morgan1933,osawa1965}. In high-resolution spectra, rare earth elements such as La, Pr, Nd, Tb and Ho may be observed to have abundances up to one million times that seen in the Sun \citep[e.g.,][]{ryabchikova2004,luftinger2010}.

The rapid oscillations in Ap stars have been studied extensively with photometry since their discovery \citep[e.g.,][]{kurtz1982,1991MNRAS.250..666M,2002MNRAS.330..153H,2016A&A...590A.116J,2018MNRAS.473...91H,cunha2019,2019MNRAS.488...18H}. The pulsations are thought to be driven by the $\kappa$-mechanism acting on the H ionisation zone \citep{2001MNRAS.323..362B}, with turbulent pressure playing a role in the excitation of some modes \citep{2013MNRAS.436.1639C}.

The pulsation modes in the roAp stars are low-degree $(\ell\leq3)$, high-overtone $(n\gtrsim15)$ asymmetric modes that have periods in the range $4.7-23.7$\,min. The pulsation axis is closely aligned to the magnetic axis which is in turn misaligned with the rotation axis. This leads to oblique pulsation, resulting in a varying pulsation amplitude over the rotation cycle of the star as the observer sees the pulsation pole at different aspects. This oblique pulsator model was formulated by \citet{kurtz1982} and has subsequently been improved in many works \citep[e.g.,][]{1985PASJ...37..245S,1985PASJ...37..601S,2002A&A...391..235B}.

Despite photometric observations of roAp stars dominating the literature, time-resolved spectroscopic observations have the ability to provide many more detailed insights into the physics of Ap and roAp stars. New insights from radial velocity studies of spectral lines in the roAp stars were provided by \citet{1999AstL...25..802S} through the analysis of 14 nights of high-resolution ($R\sim35\,000$) observations of $\gamma$\,Equ, a bright well-studied roAp star. They found that pulsation amplitudes were highest in lines of doubly ionised Pr and Nd, while lines of Ba\,{\sc{ii}} and Fe\,{\sc{ii}} were essentially stable. \citet{2002A&A...384..545R} showed that, in the case of $\gamma$\,Equ, elements such as Ba and Fe were located low in the stellar atmosphere, below $\log\tau_{5000}=-1.0$, whereas Pr and Nd could be found much higher in the atmosphere in regions where $\log\tau_{5000}=-8.0$ \citep[although this was later refined by][to be in regions where $\log\tau_{5000}=-5.0$]{2005A&A...441..309M}. This showed that the pulsations in roAp stars are a strong function of atmospheric depth which can be studied through the analysis of spectral lines of different elements, and with line bisectors \citep[e.g.,][]{2008MNRAS.386..481E}.

Furthermore, time-resolved spectroscopic observations of Ap stars have the ability to detect low amplitude modes of only a few \ms\ which were not detected in ground-based photometric observations \citep[e.g.,][]{2013MNRAS.431.2808K}. Even with the advent of space-based photometric survey missions such as the Transiting Exoplanet Survey Satellite \citep[TESS;][]{2015JATIS...1a4003R}, some roAp stars still only show variability in spectroscopy and not in the $\upmu$mag precision photometric data. 

This increased sensitivity to pulsation is important in two ways. Firstly, it provides the opportunity to robustly determine if an Ap star belongs to the roAp group or is a non-oscillating Ap (noAp) star. \citet{2002MNRAS.333...47C} calculated a theoretical instability strip for the roAp stars and, when plotting this and the positions of the known roAp stars on an HR diagram, found that many of the class members are cooler than the red-edge of the theoretical instability strip. Furthermore, there were no stars close to the hot blue-edge. It is still not known if this is an observational bias induced from target selection or the sensitivity of observations. Although TESS observations will help to answer this question \citep{cunha2019}, spectroscopy is still the more powerful tool to fully probe the extent of the pulsational behaviour in the roAp stars. Secondly, with the ability to detect low-amplitude modes, time-resolved spectroscopic observations allow for a more complete asteroseismic solution of a pulsating star. The presence of a regular pattern of modes in the roAp stars, i.e. when the modes are in the asymptotic regime \citep{1980ApJS...43..469T,1990ApJ...358..313T}, provides the opportunity for a more precise model of the star to be constructed, allowing the interactions between pulsation, rotation, magnetic fields, and chemical stratification to be fully investigated.

To that end, we present the first successful use of the High Resolution Spectrograph (HRS) mounted on the 10.0-m Southern African Large Telescope (SALT) to detect high-frequency pulsations in a well studied roAp star, $\alpha$\,Circini. This study is motivated by the fact that TESS is, and will continue to be, observing almost 1400 known Ap stars in 2-min cadence for 27\,d or more in the search of pulsations. These data, achieving $\upmu$mag precision, are increasing the number of known roAp stars. However, the TESS photometric data are not as sensitive to low-amplitude modes as spectroscopy is, so ground-based time-resolved spectra are still needed to fully exploit these new class members. By testing the capabilities of the SALT HRS instrumentation, we are proving the suitability of another resource for obtaining these much-needed observations.

\section{The target}
$\alpha$\,Circini is the brightest known roAp star ($V=3.2$). Frequency analysis of photometric data by \citet{kurtz1994} showed $\alpha$\,Cir to pulsate with a dominant frequency of 211\,d$^{-1}$ (2442\,$\upmu$Hz, $P=6.8$\,min), with several other low amplitude frequencies present. Their multisite campaign enabled them to deduce a rotation period of $4.48$\,d through identification of rotationally split side lobes to the principal peak. Further to deriving the period, they noted that over the rotation cycle the amplitude of the pulsation was only slightly modulated, an effect of the oblique pulsation. This result shows that the amplitude of the pulsation in $\alpha$\,Cir does not go to zero at quadrature, thus observations at any rotation phase will allow for the detection of the pulsation(s).

Further to extensive photometric observations, $\alpha$\,Cir has been the subject of several high-resolution spectroscopic studies \citep[e.g.,][]{1998MNRAS.295...33B,2001A&A...377L..22K,2003MNRAS.344..242B}. In particular, \citet{2006MNRAS.370.1274K} observed the star with the Ultraviolet and Visual Echelle Spectrograph (UVES) on the Very Large Telescope (VLT), while \citet{2013ASPC..479..115M} presented early results from the High Accuracy Radial velocity Planet Searcher (HARPS) spectrograph on the La Silla 3.6-m telescope. \citeauthor{2006MNRAS.370.1274K} observed $\alpha$\,Cir for 2\,hr at a cadence of 26.5\,s resulting in a data set of 265 spectra. Their analysis showed that the principal pulsation mode had an amplitude of almost $1\,000$\,m\,s$^{-1}$, with two lower amplitude peaks at 192 and 59\,m\,s$^{-1}$. The amplitude ratio between the first and second peaks was 3.5 times greater than that measured via ground-based photometry, demonstrating the power of spectroscopy. With these data, the authors showed that the amplitudes of pulsation in the rare earth lines, which form high in the atmosphere, are greater than that seen in the core of H$_\alpha$ that forms closer to the continuum level where photometry is most sensitive. 

The HARPS spectra used as a comparison in this work are a subset of over 4\,800 spectra obtained on nine nights in 2008 February and April. An initial analysis of the full data set showed 36 periodic signals with amplitudes from 52\,\ms\ down to an amplitude of 56\,cm\,s$^{-1}$, of which 30 of these frequencies were previously unknown \citep{2013ASPC..479..115M}. The analysis of the HARPS data used `integral' RV measurements which, although providing higher precision, remove information about individual lines, and thus atmospheric height \citep{2004MNRAS.351..663H}. 

\section{Observations and data treatment}

With the need to use high efficiency telescope-instrument combinations for the study of short period pulsations (making the exposure time a small faction of the pulsation period to reduce apodization), we tested the capabilities of the Southern African Large Telescope \citep[SALT;][]{buckley2006} and its High Resolution Spectrograph \citep[HRS;][]{bramall2010,crause2014}. SALT is a 10-m class telescope consisting of 91 1-m hexagonal mirror segments. The telescope has a fixed altitude pointing and is steerable in the azimutal direction only. This configuration results in an annulus on the sky where the telescope can observe at a given time, limiting the duration that a target can be observed for. The instruments, or fibre heads, are mounted on a tracker over the primary mirror allowing the light to be fed to the instrument over the course of the observations. The HRS is a fibre-fed, dual-beam, echelle spectrograph with wavelength coverage of $3700-5500$\,\AA\ and $5500-8900$\,\AA\ in the blue and red arms, respectively. 

In this section we discuss the HRS observations and data processing, as well as a subset of the aforementioned HARPS data set that we used to compare and evaluate the capabilities of HRS.

\subsection{HRS data}

We observed \alpcir\ with SALT HRS in its High Stability (HS) mode, with a resolving power of $R\sim65\,000$, using 8-s exposures. We used non-standard readout modes, opting for {\sc{fast}} readout and the use of multiple amplifiers for each chip (two for the blue arm, 4 for the red arm). This is the first time this mode has been exploited for scientific investigation. This mode reduced the readout and saved time from the standard 60\,s to 10\,s, increasing the cadence of observations and enabling us to record an extra 105 spectra in the observing window. The data were acquired on 2017 May 14 in photometric conditions with 1.5\,arcsec seeing, and cover a period of 2.45\,h with a break of 10.23\,min due to an instrument fault. In total, 279 spectra were recorded. Given the non-standard set up of the observations, a bespoke version of the P{\sc{y}}HRS code \citep{crawford2016} was used to extract the 1D spectra for each arm. The data underwent standard calibrations, including flat-fielding, order extraction, and wavelength calibration using a ThAr arc. 

Fully-reduced 1D spectra from each arm were independently normalised with low-order polynomials, manually continuum fitted, and bad chip regions were removed. After removal of cosmic rays, the two spectral regions were merged.

Upon inspection, 31 spectra after the instrument fault were affected by a variable CCD temperature rendering them unsuitable for scientific exploitation and were removed from the data set. This resulted in a final total of 248 usable spectra covering a period of 2.45\,h with a 26.03\,min break in the data.

Since the observed pulsation amplitude in \alpcir\ is variable over the stellar rotation period, it is useful to know at what rotation phase the observations were obtained. To determine this, we took the rotation period to be $4.4790\pm0.0001$\,d \citep{kurtz1994} and a zero-point in time as the pulsation maximum given in table\,2 of \citet{kurtz1994}. We calculated that our HRS observations were obtained at rotation phase $\phi_{\rm rot}=0.32\pm0.20$.

\subsection{HARPS data}
Since our aim is to benchmark HRS against other high resolution spectrographs used for asteroseismology of roAp stars, we have extracted data from the European Southern Observatory (ESO) archive of \alpcir\ taken with the HARPS spectrograph ($R\sim 115\,000$). Since the HRS data cover just 2.45\,hr, we selected a subset of spectra from the HARPS data set to conduct the analysis. The pulsation amplitudes in roAp stars can vary over the rotation period as a result of oblique pulsation. To minimise this effect between the two data sets, we selected a subset of HARPS data as close as possible in rotation phase to those obtained with HRS. Again, using the rotation period of $4.4790\pm0.0001$\,d from \citet{kurtz1994}, and now a zero-point in time as the {\it middle} of the HRS observations (BJD=2457887.499902), we selected data from the HARPS observations that were obtained at a rotation phase difference of $\Delta\phi_{\rm rot}=0.04\pm0.07$. Within $1\,\sigma$, the observations cover the same rotation phase in the two data sets, despite being obtained 3315\,d apart. This subset of HARPS data consisted of 153 spectra each with an exposure time of 20\,s, covering 2.45\,hr. We have used fully-reduced data publicly available from the ESO data archive for the analysis.

As with the HRS data, the HARPS 1D spectra from the red and blue spectral regions were independently normalised with low-order polynomials, manually continuum fitted then stitched together with unity values inserted over a chip gap. Cosmic rays were removed.

\subsection{Cross-Correlation \& Radial Velocity Measurement}
 We visually selected spectral lines in the HARPS spectra suitable for cross-correlation. Criteria for selection were that the line was isolated (in the core region), free from telluric line interference or contamination from Balmer line wings, and showed significant visual standard deviation with a two-bump shape across the observations (cf. Fig.\,\ref{fig:LPV}). The two-bump structure originates from the primary pulsation frequency at $210$\,\cd\ (section \ref{lpv}). Lines were identified from line lists for HD\,101065\footnote{\url{http://www-personal.umich.edu/~cowley/prznew2.html}}, 10\,Aql\footnote{\url{http://www-personal.umich.edu/~cowley/10Aql.html}}, and were supplemented with lines from \citet{2006A&A...456..329R}, \citet{2007A&A...473..907R} and \citet{bruntt2008}.
 
 These lines were cross-matched with the HRS spectra and any not appearing in the latter were removed. In total, 49 individual lines of 9 elements including the core of H$_\alpha$ were identified for analysis. Our line list was dominated by 28 lines of Nd\,{\sc{iii}}, with the full list of the lines used for each element given in Table\,\ref{tab:line_lists}.

\begin{table}
\caption{Spectral lines of each element and ionisation state that were used in the cross-correlation analysis. We have used the same lines for the analysis of the HRS and HARPS data sets to provide a fair comparison.}
\centering
\begin{tabular}{llllll}
\hline
Element & N$^{\rm o}$ of & \multicolumn{2}{l}{Wavelength}\\
    /Ion       & lines & (\AA)\\
\hline
    H$_\alpha$ & 1 & 6562.79 \\
\hline
Ce\,{\sc{ii}} & 3 & 4071.81& 4073.48& 4628.16 \\
\hline
Pr\,{\sc{iii}} & 7 & 4713.70& 5284.70& 5299.99& 5956.05 \\
               &   & 6090.02& 6160.24& 6195.63 \\
\hline
Nd\,{\sc{ii}} & 5 & 4021.69& 4061.09& 4903.24& 4942.97 \\
                & & 5293.16\\
\hline
Nd\,{\sc{iii}}& 28 & 4570.63& 4624.98& 4627.26& 4711.33 \\
              &    & 4759.55& 4769.62& 4796.51& 4810.37 \\
              &    & 4911.66& 4912.94& 4914.10& 4927.48 \\
              &    & 5050.69& 5102.42& 5127.04& 5151.73 \\
              &    & 5193.04& 5203.92& 5264.96& 5286.75 \\
              &    & 5677.18& 5802.53& 5845.02& 5851.54 \\
              &    & 5987.68& 6145.07& 6327.26& 6550.23 \\
\hline
Sm\,{\sc{ii}} & 2 & 4420.53& 4424.34 \\
\hline
Tb\,{\sc{iii}}& 1 & 5505.41 \\
\hline
Dy\,{\sc{iii}}& 1 & 4510.03 \\
\hline
Ho\,{\sc{iii}} & 1 & 4494.52 \\
\hline
\end{tabular}
\label{tab:line_lists}
\end{table}

The central wavelength and equivalent width (0$^{\rm th}$ moment) of each line in an element/ion set was measured by the moment method \citep{1992AandA...266..294A}. These were constructed into a $\delta$-function template using the wavelengths as positions and the equivalent widths as the depth to weight the lines \citep[see][]{NewEntry3}. Cross-correlation with this template produced line profiles that combine the information of the variation from all the lines chosen. Cross-correlation was done for different ionisation states for each element as different ions for the same element can have differing phases and amplitudes in roAp stars due to the stratification of elements and the presence of false nodes between line forming layers in the atmosphere \citep{2018MNRAS.480.1676Q}. From the cross-correlation profiles the radial velocity measurements were made using the first moment \citep{1992AandA...266..294A}.

\section{Radial Velocity analysis}

With each set of radial velocity (RV) measurements, we pre-whitened the RV curve in a iterative way to remove low-frequency variations caused by either stellar rotation or instrumental artefacts. This procedure was performed in the frequency range $0-30$\,\cd, to an amplitude limit that was determined from the noise peaks at high frequency. For the HRS data that limit was in the range $10-66$\,m\,s$^{-1}$, while the range for the HARPS data was lower ($11-42$\,m\,s$^{-1}$). This noise difference is a result of a combination of integration time, efficiency and the different resolution of each spectrograph.

\begin{figure*}
\centering
\includegraphics[width=0.49\textwidth]{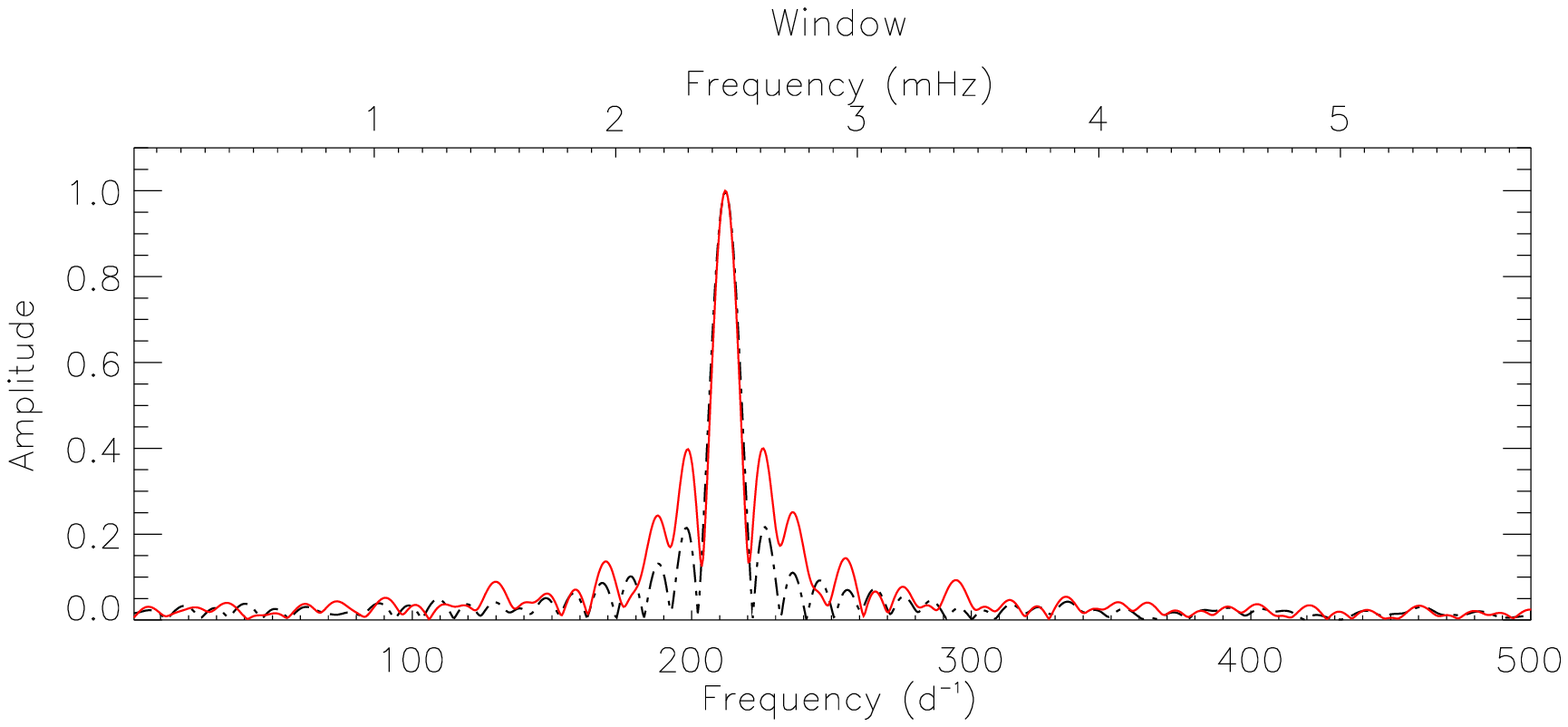}
\includegraphics[width=0.49\textwidth]{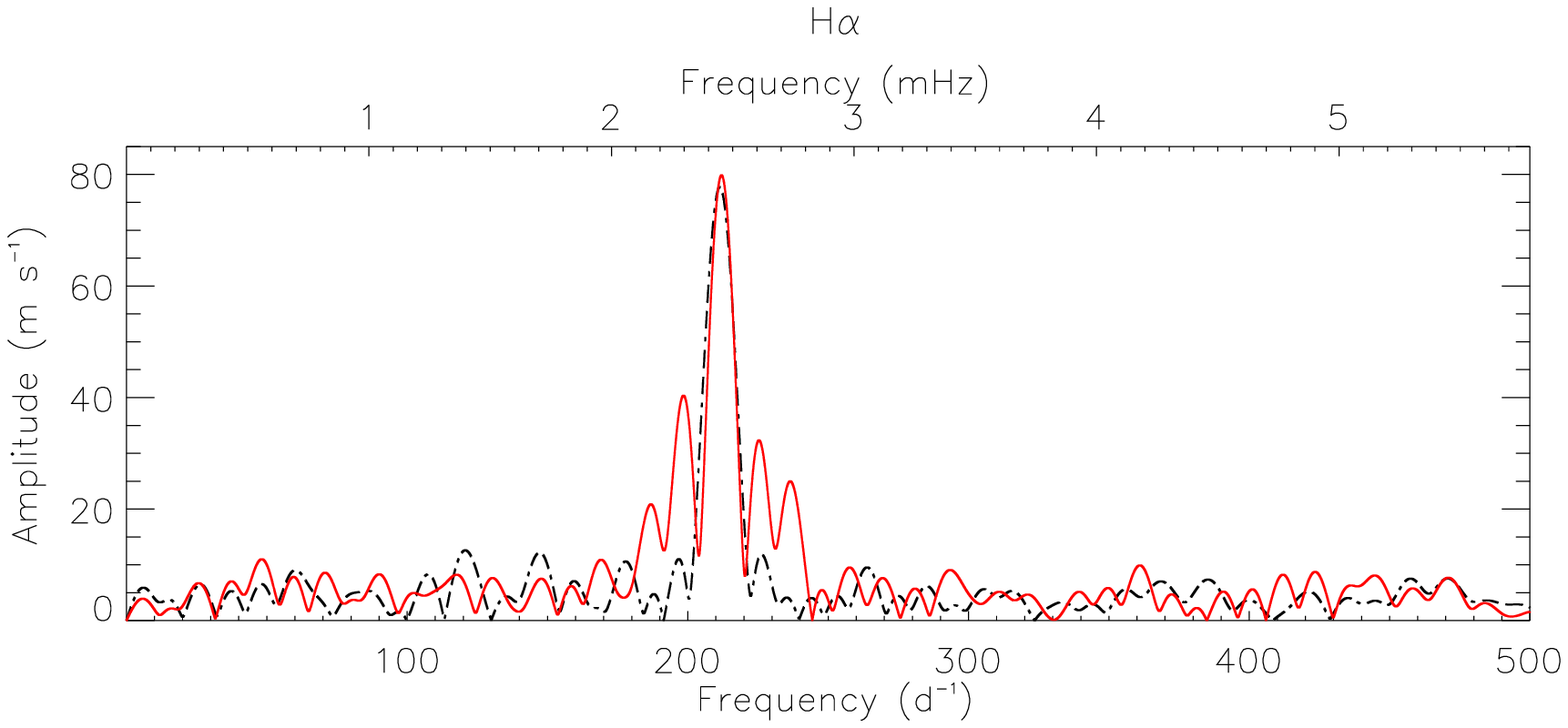}
\includegraphics[width=0.49\textwidth]{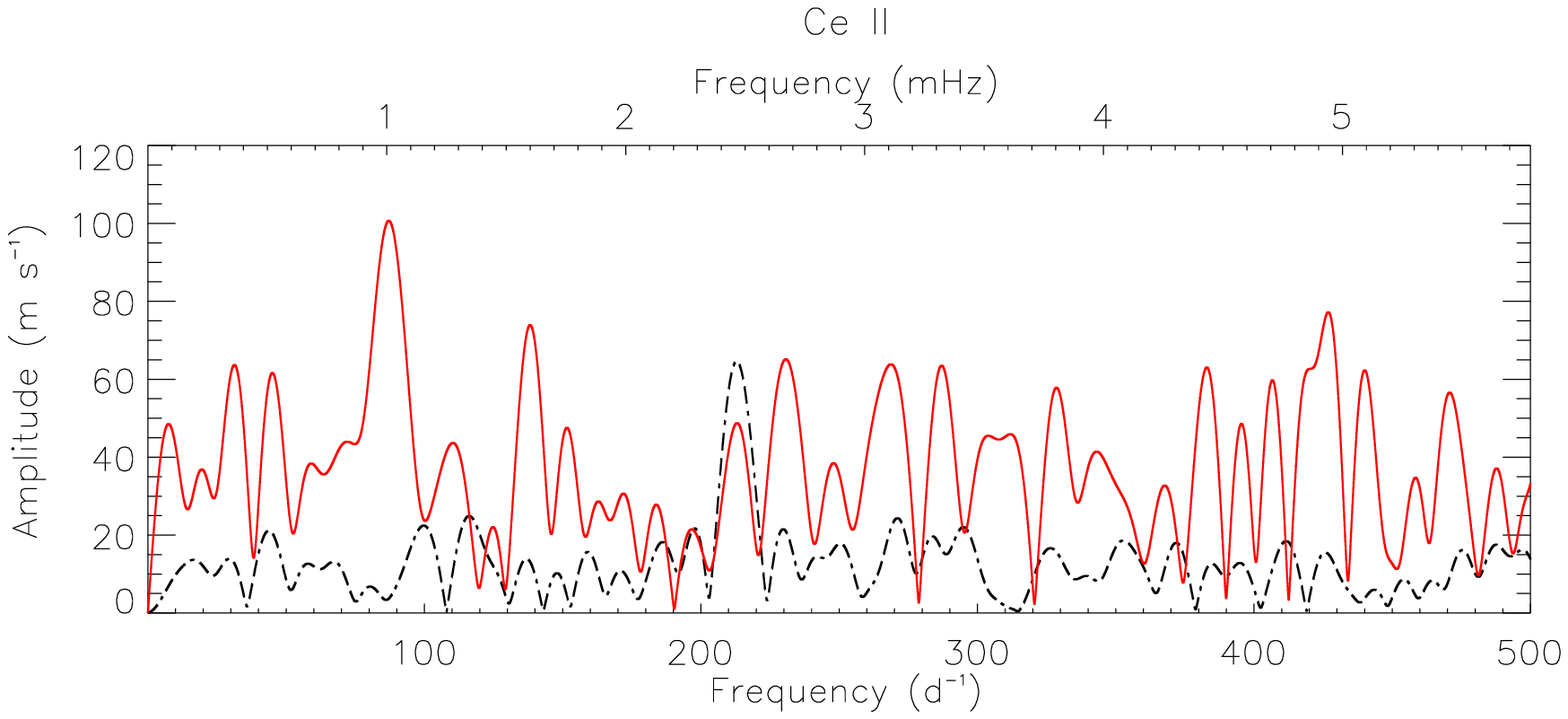}
\includegraphics[width=0.49\textwidth]{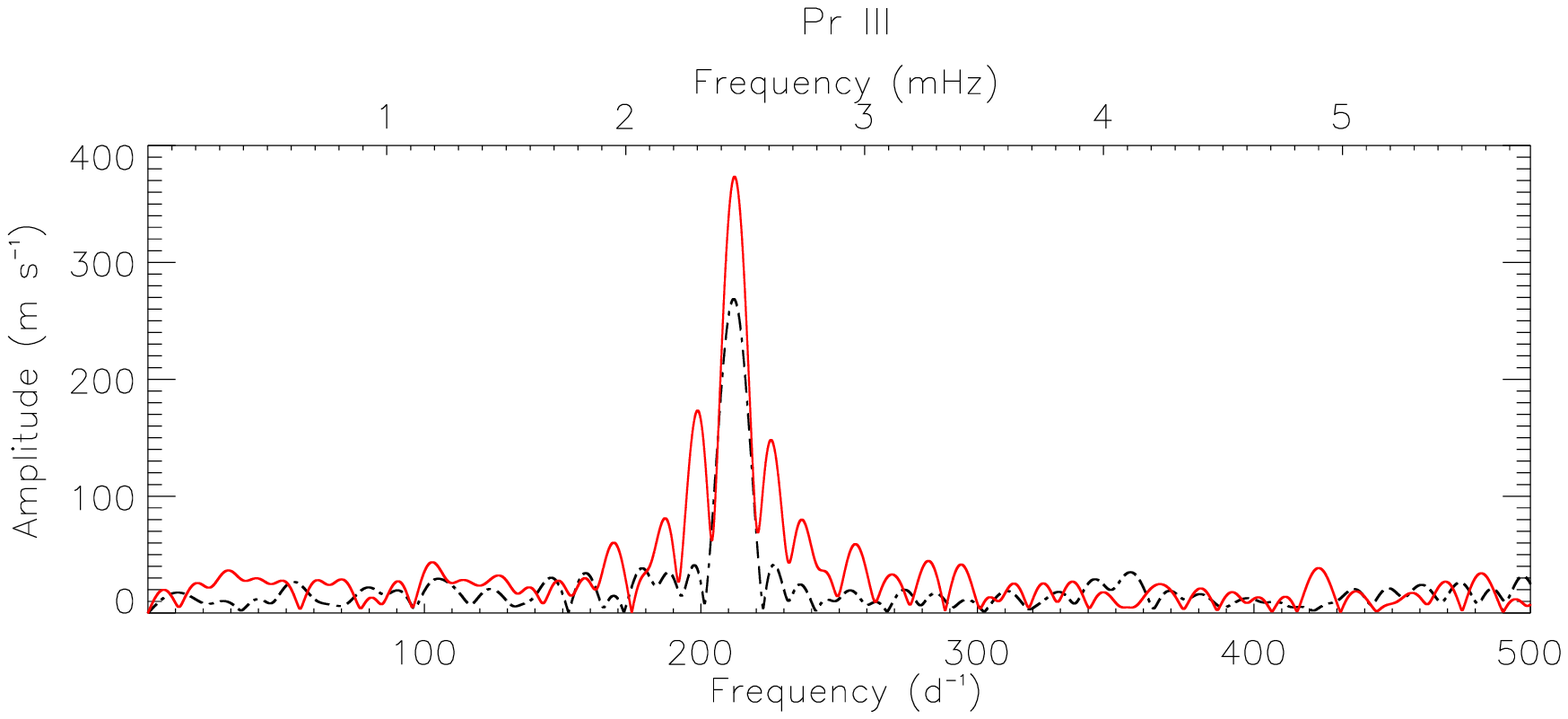}
\includegraphics[width=0.49\textwidth]{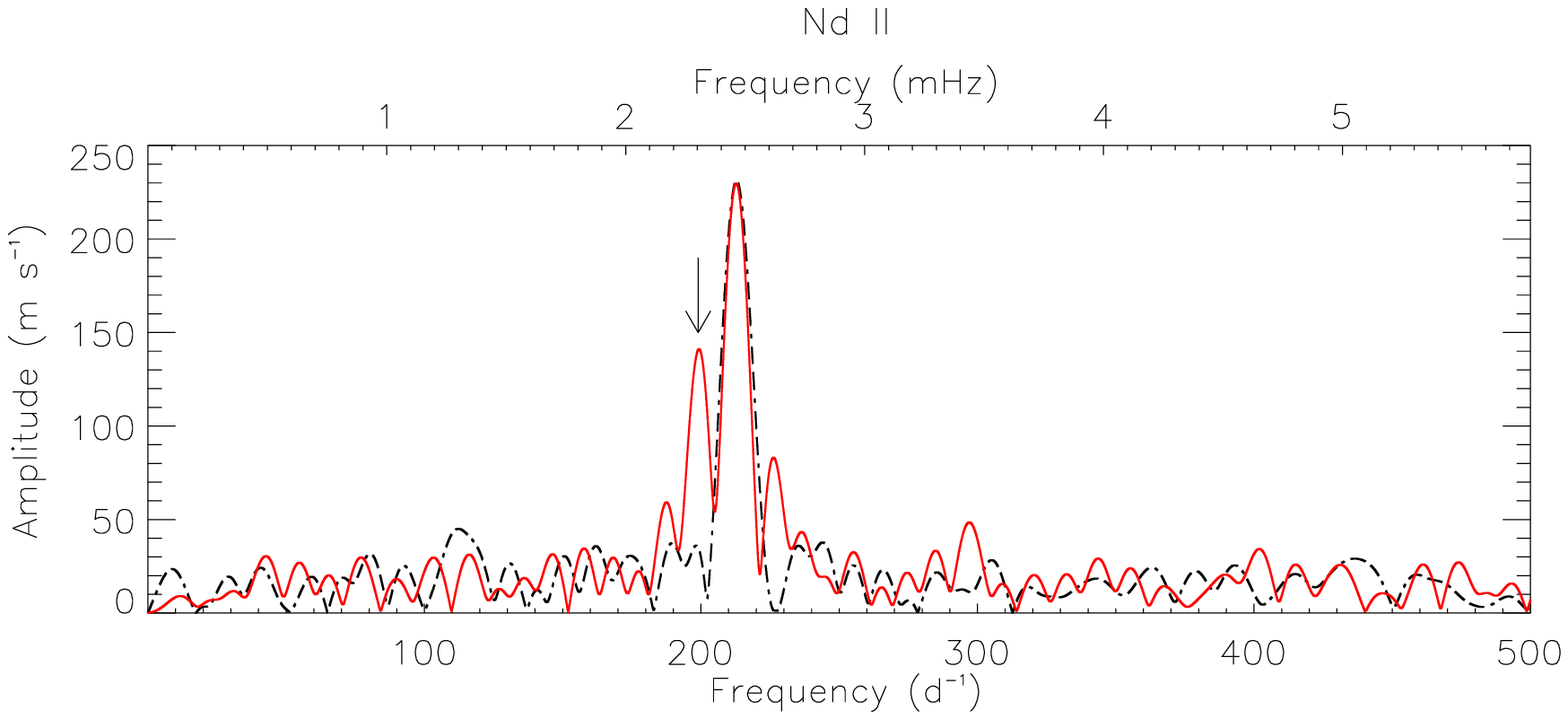}
\includegraphics[width=0.49\textwidth]{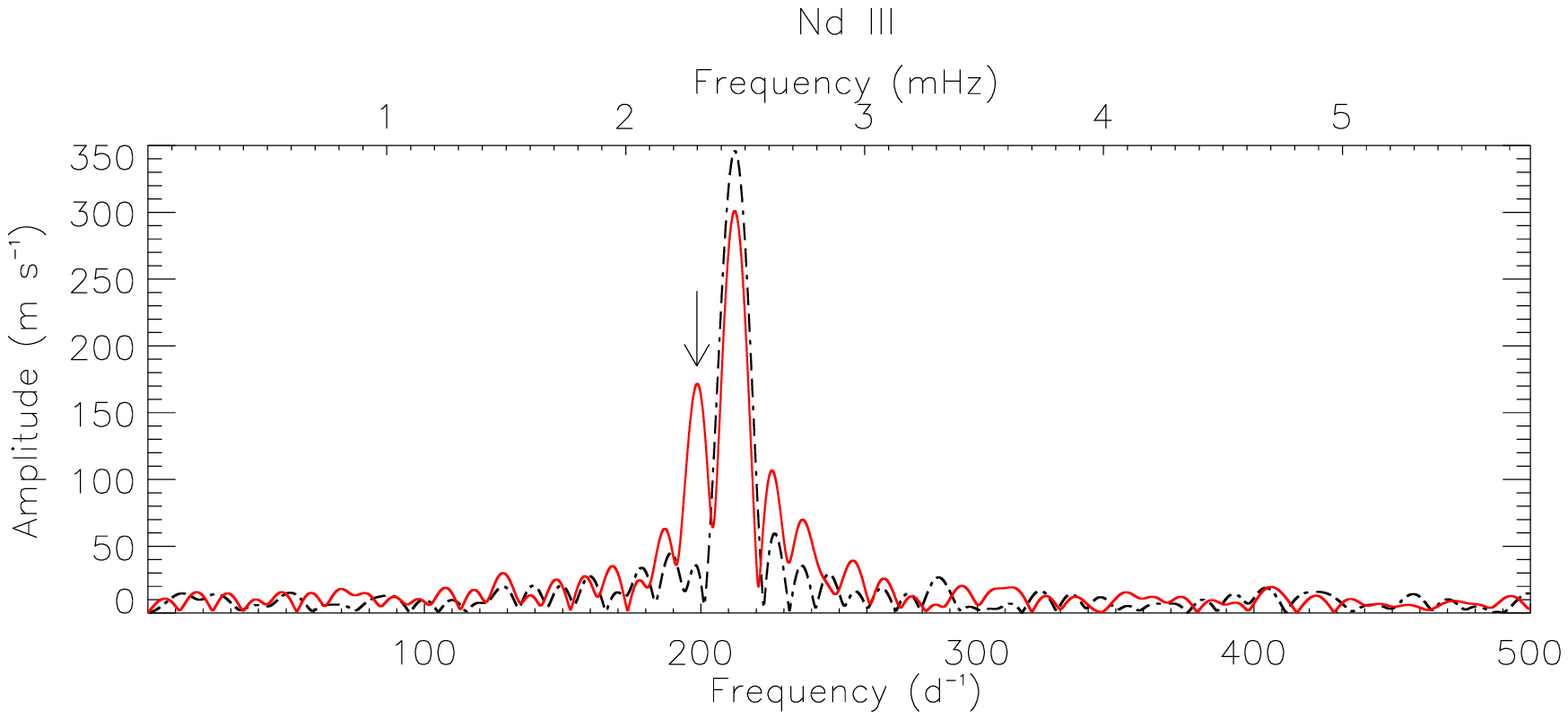}
\caption{Comparison of the amplitude spectra of the HRS RV data (red solid line) with the HARPS RV data (black dashed line). The window function for each data set is shown in the top left plot, with the amplitude normalised to unity. Note the changing amplitude scale from plot to plot. The lower frequency sidelobe of the HRS observations of Nd\,{\sc{ii}} and Nd\,{\sc{iii}} lines are not entirely reminiscent of the window function (nor the HARPS data) and are showing the contribution of a second pulsation mode at about 200\,\cd\ (2.3\,mHz), as indicated by the arrow. The other elemental ions considered in Table\,\ref{tab:line_lists} are shown in Appexdix\,\ref{sec:appendix}.}
\label{fig:RV_ft}
\end{figure*}

For each set of radial velocity measurements, we calculated the amplitude spectrum in the range $0-500$\,\cd\ using the discrete Fourier transform method of \citet{1985MNRAS.213..773K}. This frequency range covers the known pulsation frequencies and the harmonic of the dominant mode. We present the results in Fig.\,\ref{fig:RV_ft} where the red heavy lines are the HRS results and the black dashed lines are the HARPS results. The top left panel shows the window function of the two data sets.

The results clearly demonstrate that SALT HRS is very capable of detecting pulsation in Ap stars. There are two obvious differences between the results of the two data sets: the window function and the asymmetric side lobes of the HRS Nd\,{\sc{ii}} and Nd\,{\sc{iii}} spectra. The non-sinc function appearance of the HRS window function results from the 26\,min gap in the data caused by an instrument fault. Although not ideal, this gap has allowed us to show that even with breaks in the observations, HRS is capable of detecting pulsations. The asymmetric side lobes of the Nd lines in the HRS data is most likely caused by the presence of a second mode at about 200\,\cd\ that will be discussed later.

To provide a quantitative comparison of the two data sets, we have fitted the following function by non-linear least squares to each of the element/ion results after removing the mean RV:
\begin{equation}
\Delta \rm{RV} = A \cos (2\pi\nu(t - t_0) + \phi)\,,
\label{eq:fn_fit}
\end{equation}
where $A$ is the semi-amplitude, $t$ is the time of observation, $t_0$ is a reference time (taken to be the midpoint of the observations independently for each data set), and $\phi$ in the phase. The results are shown in Table\,\ref{tab:all_nlls}, demonstrating that the frequency for each element/ion agree to within $1.8\,\sigma$ between the two data sets. The phase results can be compared within in each data set, but not {\it between} data sets as the zero-point is different.

\begin{table*}
\caption{Results of a nonlinear least-squares analysis of the RV data from HARPS and HRS, respectively. The phases are with respect to the midpoint of each data set: BJD\,2454572.88278 for the HARPS data and BJD\,2457887.49990 for the HRS data. The phase information cannot be compared between the two data sets.}
\label{tab:all_nlls}
\centering
\begin{tabular}{lcrrrcrrr}
\hline
&\multicolumn{4}{c}{HARPS} & \multicolumn{4}{c}{HRS} \\
\hline
\multicolumn{1}{c}{Element}	&	Frequency	&			\multicolumn{1}{c}{Amplitude}	&			\multicolumn{1}{c}{Phase}	&			\multicolumn{1}{c}{S/N}	&	Frequency	&			\multicolumn{1}{c}{Amplitude}	&			\multicolumn{1}{c}{Phase}	&			\multicolumn{1}{c}{S/N}	\\
	&	(\cd)	&			\multicolumn{1}{c}{(m\,s$^{-1}$)}	&			\multicolumn{1}{c}{(rad)}	&				&	(\cd)	&			\multicolumn{1}{c}{(m\,s$^{-1}$)}	&			\multicolumn{1}{c}{(rad)}	&			\\
	\hline
H\,$_\alpha$	& $211.46\pm0.22$   & $78.0\pm 3.2$     & $0.35\pm0.04$ & 24.5  & $	211.97\pm 0.19$ & $	80.1\pm	3.1 $   & $	-1.90\pm 0.04	$ &	25.8    \\
Ce\,{\sc{ii}}	& $212.89\pm0.78$   & $65.0\pm 9.6$     & $0.27\pm0.15$ & 6.8   & $	213.09\pm 3.17^*$ & $	49.0\pm	31.0 $  & $	-2.06\pm 0.63	$ &	1.6     \\
Pr\,{\sc{iii}}	& $211.85\pm0.22$   & $263.9\pm 10.3$   & $-1.68\pm0.04$ & 25.6 & $	212.14\pm 0.15$ & $	368.5\pm 11.1 $ & $	 2.35\pm 0.03	$ &	33.1    \\
Nd\,{\sc{ii}}	& $212.80\pm0.31$   & $230.0\pm 13.4$   & $-0.52\pm0.06$ & 17.2 & $	212.76\pm 0.31$ & $	230.5\pm 14.6 $ & $	-2.68\pm 0.06	$ &	15.9    \\
Nd\,{\sc{iii}}	& $212.23\pm0.11$   & $341.0\pm 6.6$    & $-0.98\pm0.02$ & 51.9 & $	212.23\pm 0.11$ & $	299.7\pm 6.5 $  & $	-3.10\pm 0.02	$ &	46.3    \\
Sm\,{\sc{ii}}	& $211.58\pm0.35$   & $172.1\pm 11.3$   & $0.15\pm0.07$ & 15.2  & $	212.46\pm 0.38$ & $	226.8\pm 17.1 $ & $	-2.00\pm 0.08	$ &	13.3    \\
Tb\,{\sc{iii}}	& $212.10\pm0.31$   & $654.8\pm 37.9$   & $-2.89\pm0.06$ & 17.3 & $	212.30\pm 0.47$ & $	550.0\pm 51.6 $ & $	 1.27\pm 0.09	$ &	10.7    \\
Dy\,{\sc{iii}}	& $212.73\pm0.38$   & $328.2\pm 23.4$   & $-0.07\pm0.07$ & 14.0 & $	212.55\pm 0.42$ & $	365.0\pm 31.0 $ & $	-2.35\pm 0.08	$ &	11.8    \\
Ho\,{\sc{iii}}	& $211.28\pm0.47$   & $163.3\pm 13.9$   & $-1.92\pm0.09$ & 11.7 & $	211.50\pm 0.71$ & $	191.7\pm 27.2 $ & $	 2.32\pm 0.14	$ &	7.1     \\
\hline
\multicolumn{9}{l}{Notes:$^*$This frequency was extracted with a search around the known frequency but would not have been detected in a}\\
\multicolumn{9}{l}{blind study.}
\end{tabular}
\end{table*}

It is clear from both the figure and table, that HRS was not able to detect, with confidence, variability in the Ce\,{\sc{ii}} lines, despite the lines showing a similar amplitude as the H$_\alpha$ core in the HARPS data. We inspected the spectra in the region of these lines and found no reason why we did not detect variability. The noise in the amplitude spectrum is higher than in most elements (except Tb\,{\sc{iii}} where only one line is used in the cross-correlation) and acts to hide the signal. This is perhaps a result of the lower resolution of HRS, or perhaps a change in the mode amplitude at this atmospheric height.   

The signal-to-noise ratio of the detections are similar in both data sets, with HRS occasionally out-performing HARPS. The errors on the pulsation amplitudes in the HRS data are also comparable to those obtained by \citet{2006MNRAS.370.1274K} who analysed UVES observations that cover a similar time span and number of observations presented here. We cannot make a direct comparison with those data, however, due to a significant difference in rotation phase, resulting in a different pulsation amplitude being observed.

\begin{figure}
\resizebox{\hsize}{!}{\includegraphics{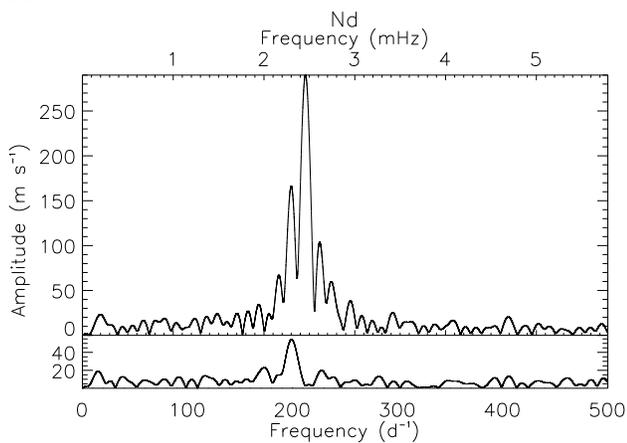}}
\caption{The principal mode and a second mode seen in the Nd lines ({\sc{ii}} and {\sc{iii}} combined). The second mode has an amplitude of $65.06\pm6.19$\,\ms\ and is found at a frequency of $199.02\pm0.63$\,\cd.}
\label{fig:Nd}
\end{figure}

As previously mentioned, there is a second peak in the Nd lines that is not seen in the HARPS data. To investigate this second peak, we created a $\delta$-function mask with both the Nd\,{\sc{ii}} and Nd\,{\sc{iii}} lines to increase precision. The amplitude spectra of the RVs for the combined Nd\,{\sc{ii}} \& {\sc{iii}} lines is shown in Fig.\,\ref{fig:Nd}. We prewhitened the RV curve to remove the principal frequency and calculated an amplitude spectrum of the residuals. This is shown in the bottom panel of Fig.\,\ref{fig:Nd}. This second frequency is found at $199.02\pm0.63$\,\cd. The frequency determination of this mode is not secure due to both its low amplitude and the length of the data set, although it is resolved from the principal mode with respect to the Rayleigh criterion. 

\begin{figure}
\resizebox{\hsize}{!}{\includegraphics{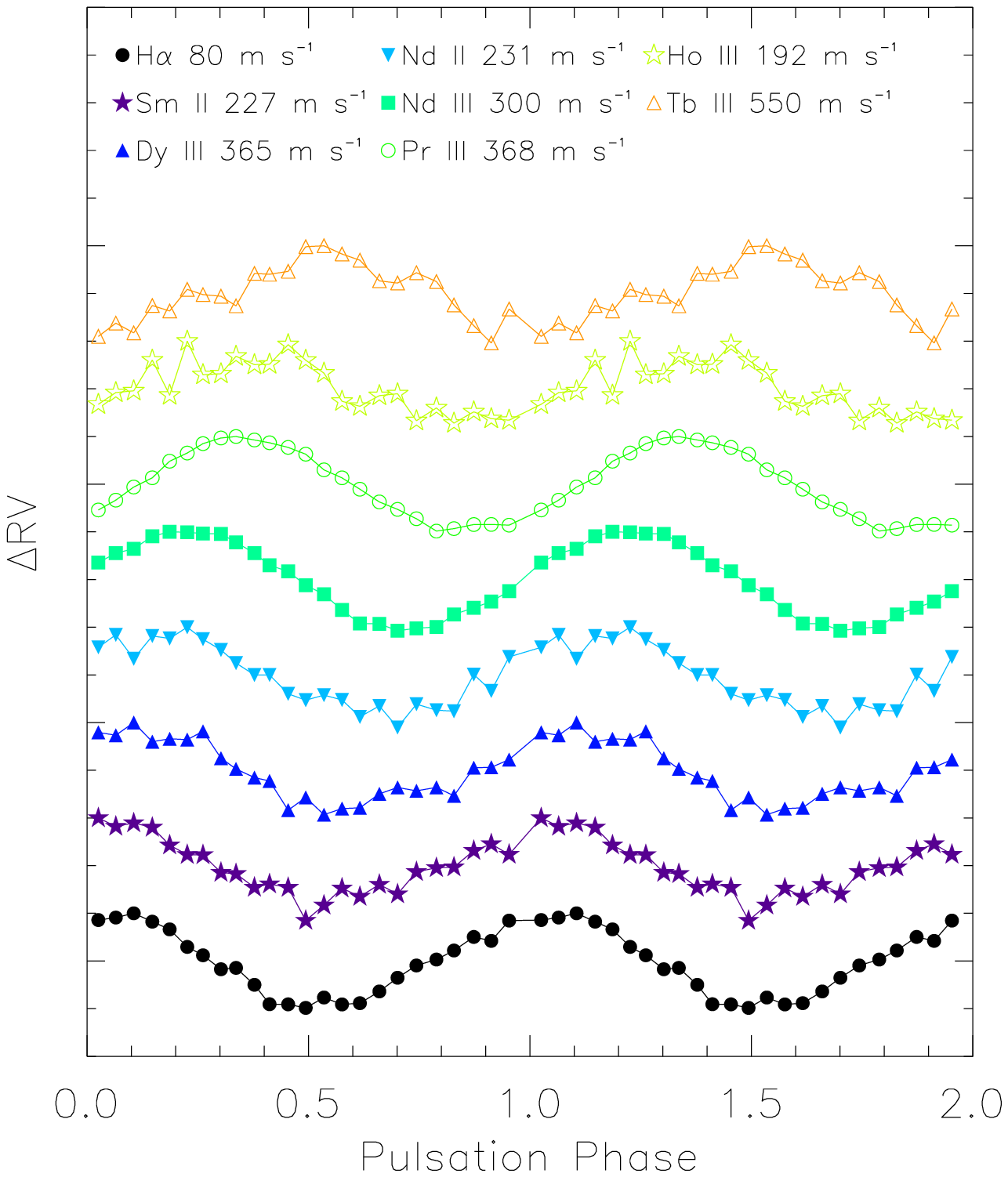}}
\caption{The RV variations for each set of analysed lines, ordered in increasing phase from H$_\alpha$. The amplitudes of the RV curves have been normalised to the maximum of each element/ion, using the amplitudes stated at the top of the figure. The variation shows how the pulsation propagates from low in the atmosphere (from H$_\alpha$) to high in the atmosphere (to Tb\,{\sc{iii}}).}
\label{fig:phased}
\end{figure}

We then investigated the phase variation of the pulsations in the different elements and ionisation states. It is well known that the elements in the atmospheres of Ap stars are stratified, meaning that as the pulsations in roAp stars propagate through the atmosphere, a phase shift should be detectable in each element and ion, indicating the relative positions of the line forming layers. We investigated this by phasing the RV curves on the pulsation period, and plotting each set of RVs in increasing pulsation phase order starting with H$_\alpha$. We show this exercise in Fig.\,\ref{fig:phased}. This figure shows how the pulsation wave propagates through the different line forming layers from $\log \tau_{5000}\sim -2$ where the H$_\alpha$ core forms, to higher than $\log \tau_{5000}\sim -5$ for the Tb\,{\sc{iii}} line forming layer, as expected in the roAp stars \citep[e.g.,][]{2007A&A...473..907R}. This changing atmospheric height also involves a change in the contribution of the acoustic and magnetic waves, with acoustic waves dominating in the upper atmosphere \citep{2018MNRAS.480.1676Q}.

The pulsation amplitudes typically follow this same pattern, increasing with increasing atmospheric height. There are two exceptions to this: the Dy\,{\sc{iii}} line that is stronger than other lines of a similar phase, and the Ho\,{\sc{iii}} line that is significantly weaker. The weakening of the Ho\,{\sc{iii}} line could be explained if it is formed close to a false node \citep{2018MNRAS.480.1676Q}, but this would not explain the amplitude seen in the Dy\,{\sc{iii}} line.

\section{Line Profile Variations and mode identification}
\label{lpv}

Spectral line profile variations (LPVs) contain rich information on the pulsation in a star and have been exploited in the study of many different types of pulsating star \citep[e.g.,][]{2003A&A...411..565N,2006A&A...452..945T,2007AN....328.1170K,2009A&A...501..291S,2018MNRAS.475.3813B}. However, in the case of the roAp stars, this is a much more complex problem due to the inclination of the pulsation axis to the rotation axis, the varying heights at which the line profiles are formed, the inhomogeneous distribution of elements, and the magnetoacoustic nature of the pulsations. Nevertheless, both theoretical \citep{2008PASJ...60...63S,2012PASJ...64....9N} and observational \citep{2006A&A...446.1051K,2007MNRAS.376..651K,2008MNRAS.386..481E} work on LPVs in roAp stars has been conducted.

\begin{figure*}
\includegraphics[width=0.49\textwidth]{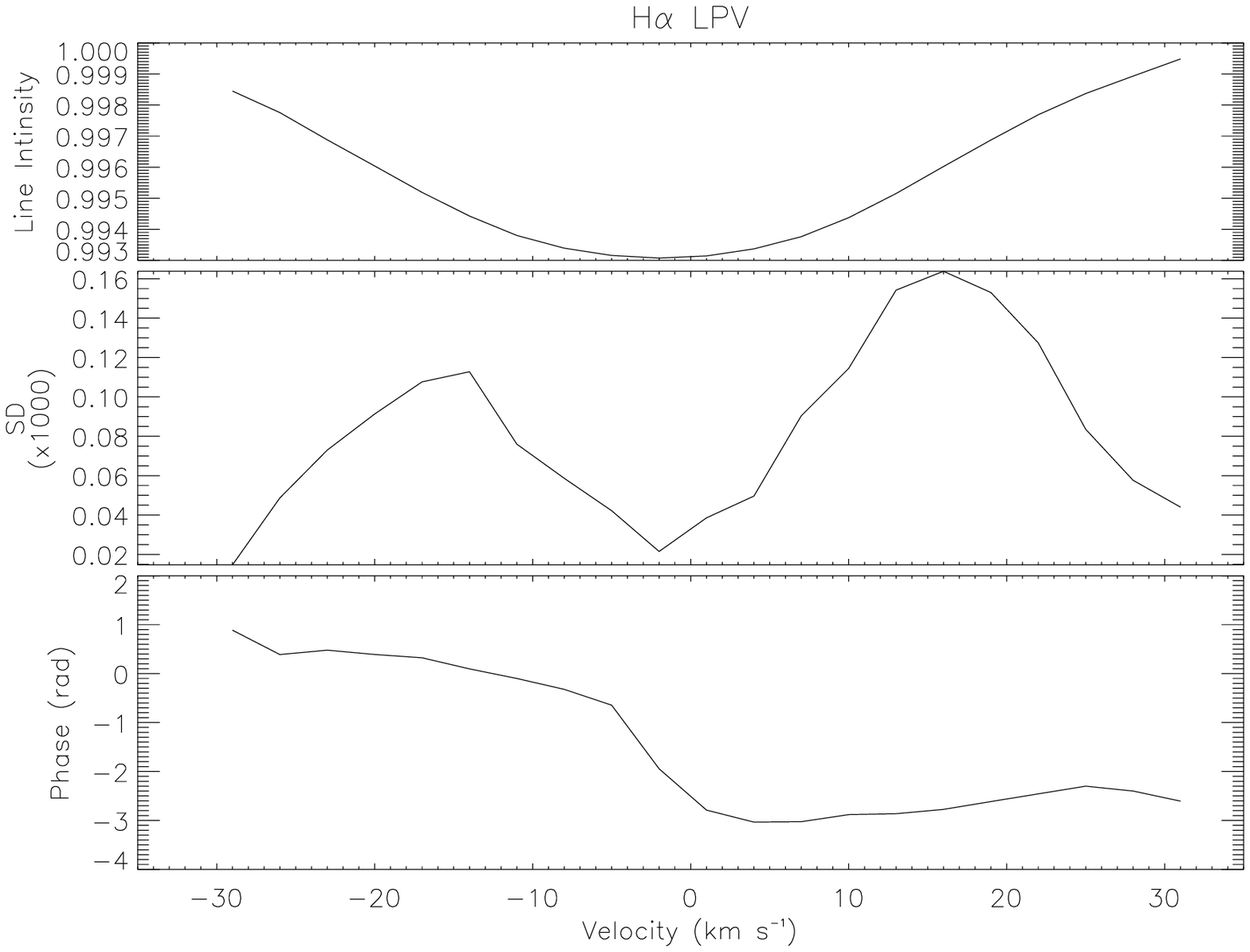}\hfill
\includegraphics[width=0.49\textwidth]{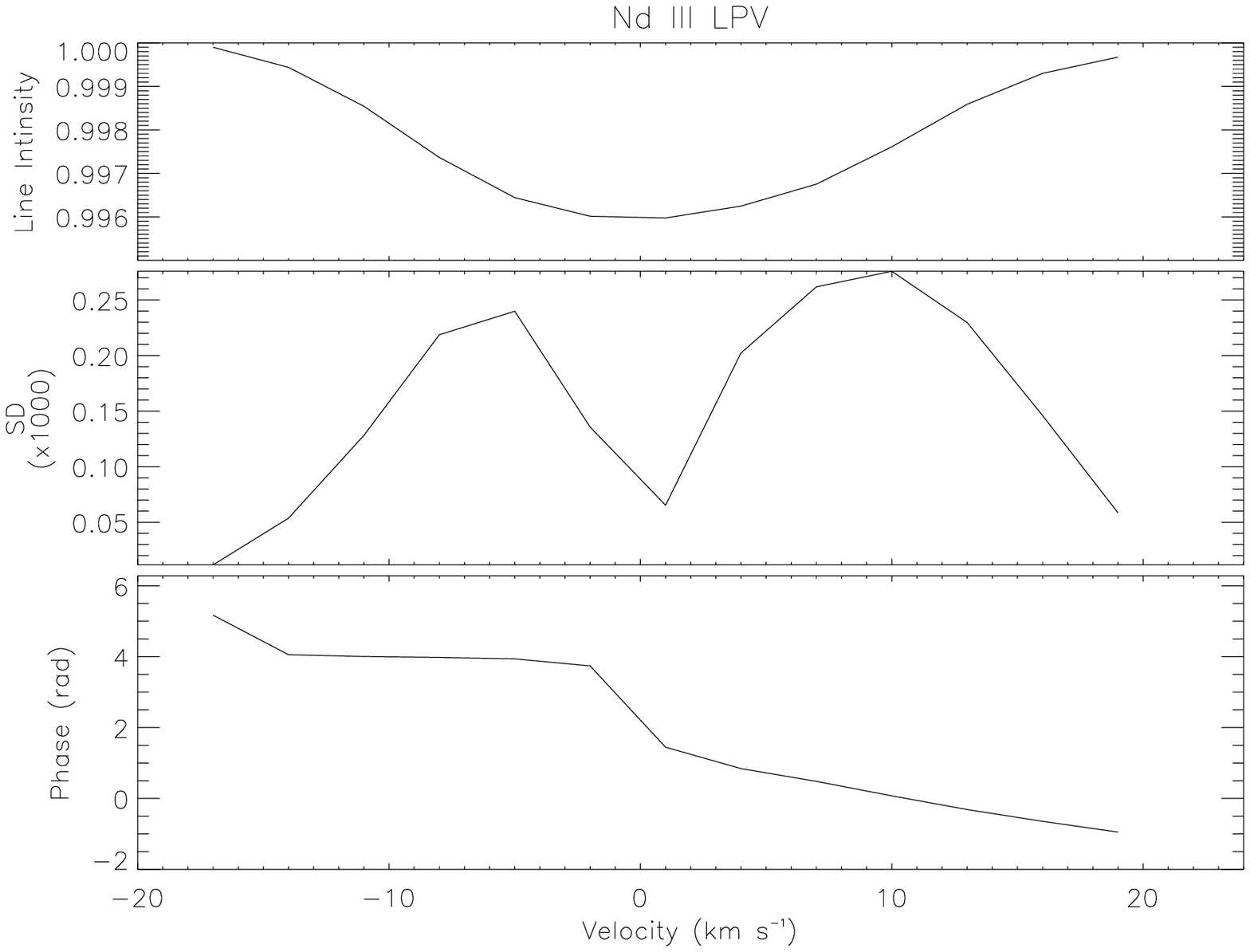}
\caption{Line profile variations of the H$_\alpha$ core (left) and the Nd\,{\sc{iii}} set of lines (right) for the HRS data. The top panel shows the mean spectral profile, the middle panel shows the amplitude of variation for a given velocity in the line, and the bottom panel shows the phase for a given velocity. In both cases, there is a phase shift of about $\pi$-rad in the line centre which is indicative of a asymmetric dipole mode. We note that there is some contribution of the second mode in the Nd\,{\sc{iii}} profile.}
\label{fig:LPV}
\end{figure*}

We are able to simplify this problem slightly in our case due to the short time span of the data set. With just 2.45\,hrs of data, our observations covered a very small portion of the rotation phase of \alpcir, allowing us to neglect the effects of rotation on the pulsation amplitudes and phases. For our short set of data, we calculated the line profile variations of the H$_\alpha$ core, and the ensemble of the 28 Nd\,{\sc{iii}} lines using the {\sc{famias}} software \citep{2008CoAst.155...17Z}. The results are shown in Fig.\,\ref{fig:LPV} with the top panel showing the average line profile, the middle panel showing the standard deviation and the bottom panel showing the change of pulsation phase across the line.

The standard deviation across the line clearly shows two symmetrical maxima in the standard deviation profile. The phase shows that the blue and red wings of the lines are varying in antiphase, as expected when the pulsation pole does not lie along the vertical plane through the line of sight (as is the case here since $\phi_{\rm rot}=0.32$).

For non-radially pulsating rotating stars, \citet{1997A&A...317..723T} derived relationships between phase variations across a line to the degree of the mode. Although our data are for an essentially non-rotating star, we used their equation\,(9):
\begin{equation}
\ell\approx 0.10 + 1.09|\Delta\phi|/\pi\,.    
\end{equation}
to estimate the degree of the mode we have detected in the HRS data. 

For H$_\alpha$, $\Delta\phi$ is about $\pi$, suggesting an $\ell=1$ mode. This is in good agreement with the many photometric observations of \alpcir. For the Nd\,{\sc{iii}} lines, $\Delta\phi$ is about 1.6$\pi$, suggesting $\ell\gtrsim2$. However, the Nd\,{\sc{iii}} lines have contributions from a second lower amplitude mode, thus adding complexity to this simple model. We draw no conclusions from this exercise, but use it to show that even a short run of HRS data can be used to constrain modes in roAp stars. This provides value added information from SALT HRS observations. With the development of more complex theoretical models of how pulsations affect LPVs in roAp stars, we expect that these, and future, HRS observations can be exploited for much more astrophysical insight.

\section{Conclusions}

We have clearly demonstrated that SALT HRS is well-suited for the detection and characterisation of pulsations in Ap stars. With just a short observing sequence, dictated by the fact that SALT has a fixed azimuth pointing, we have been able to extract two pulsation frequencies with similar S/N to those obtained with the HARPS and UVES spectrographs. We have observed how the pulsation propagates through the chemically stratified atmosphere of \alpcir, and shown that basic mode identification can be conducted using the LPVs.

We have conducted this test with the brightest member of the roAp class of variable star, and calculate these results would be achievable for stars down to a limiting magnitude of $V\sim8$, based on the S/N of individual spectra, and the HRS exposure time calculator. This may extend to fainter stars with the on-going improvements to both the telescope and instrument. 

We have shown that SALT HRS has the ability to provide the high quality observations needed for the detailed investigation of roAp stars.

\acknowledgements

DLH acknowledges financial support from the Science and Technology Facilities Council (STFC) via grant ST/M000877/1.
Some of the observations reported in this paper were obtained with the Southern African Large Telescope (SALT) under programme 2017-1-SCI-023, PI: Holdsworth. Based on observations collected at the European Organisation for Astronomical Research in the Southern Hemisphere under ESO programme 081.D-0008(A).
Some results were obtained with the software package FAMIAS developed in the framework of the FP6 European Coordination Action HELAS.
We thank the SALT team for allowing us to conduct this test. In particular we thank the SALT Astronomer, Marissia Kotze, and the SALT Operator, Veronica van Wyk, for their perseverance when taking the observations, and Steve Crawford for adapting the P{\sc{y}}HRS software to reduce the data.
We are grateful to Prof. Don Kurtz for useful comments and discussion on the manuscript, and an anonymous referee.

\appendix

\section{Amplitude spectra of remaining RV measurments}
\label{sec:appendix}
Remaining panels from Fig.\,\ref{fig:RV_ft} comparing the amplitude spectra of the RV variations from elements extracted from the HRS and HARPS data.

\begin{figure}[!h]
{\includegraphics[width=0.49\textwidth]{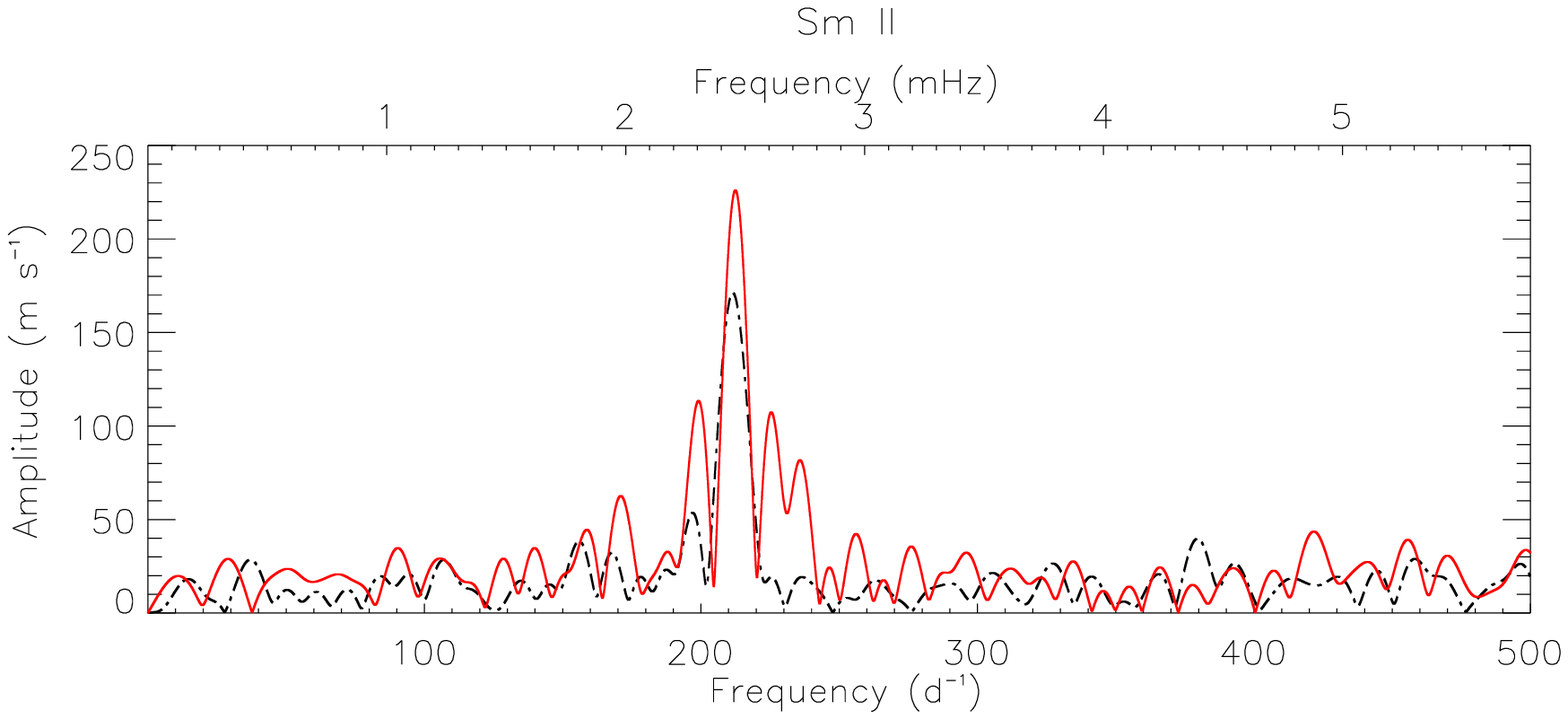}}
{\includegraphics[width=0.49\textwidth]{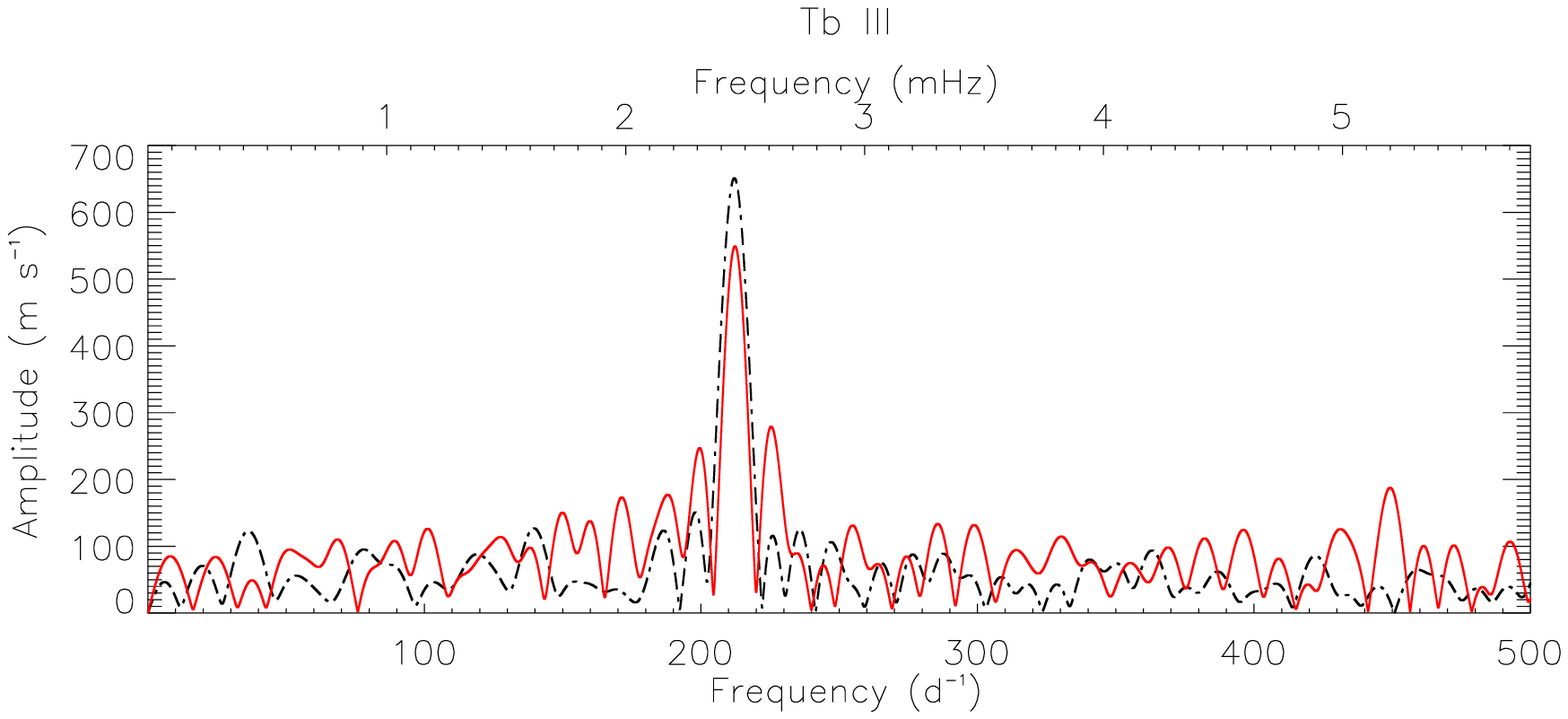}}
{\includegraphics[width=0.49\textwidth]{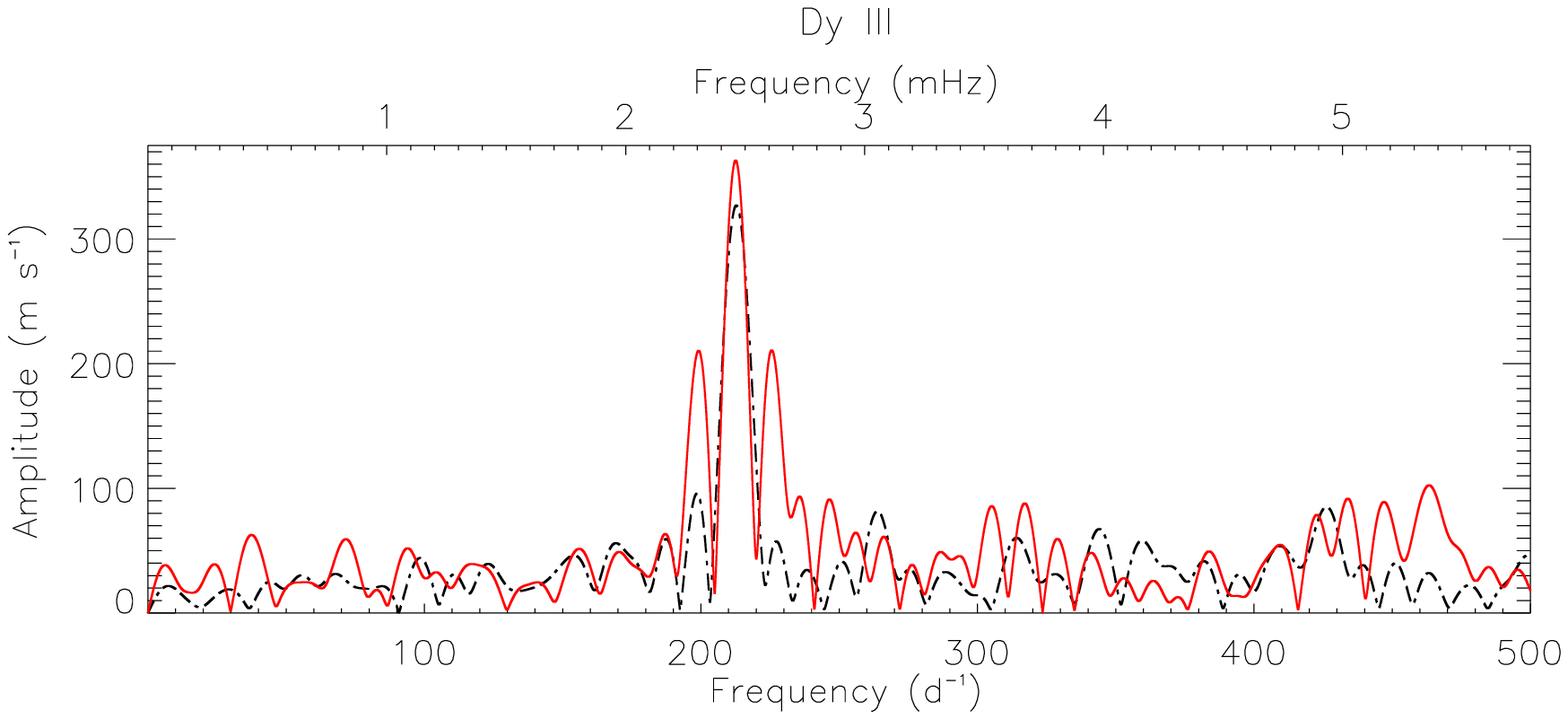}}
{\includegraphics[width=0.49\textwidth]{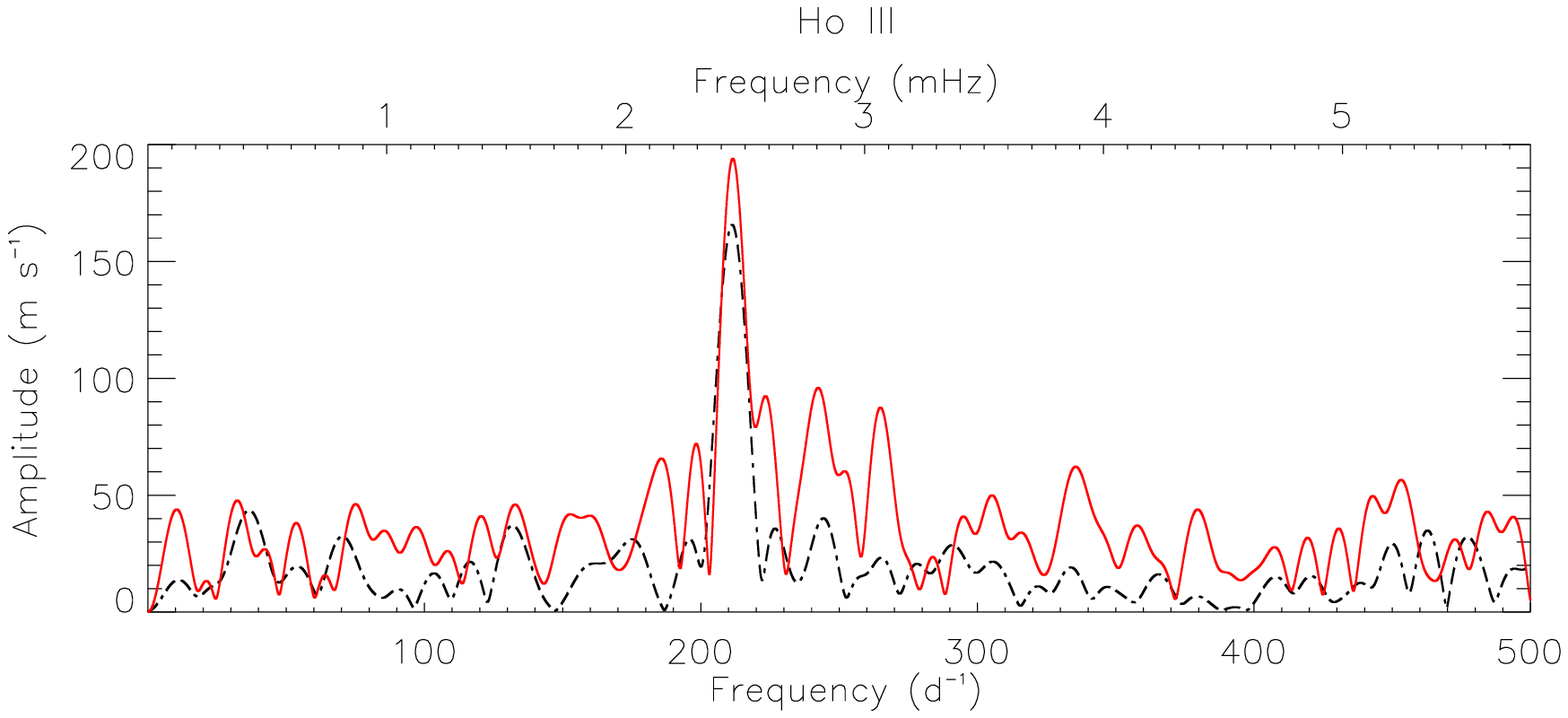}}
\caption{Extension of Fig.\,\ref{fig:RV_ft}.}
\label{fig:RV_ft_appendix}
\end{figure}

\bibliographystyle{aa}
\bibliography{alpha_cir}{}

\end{document}